\documentclass[epjST]{svjour}
\usepackage{graphics}
\usepackage{epsfig}
\usepackage{ dsfont }
\begin{document}
\title{Hamilton-Jacobi formulation of holographic Entanglement Entropy}
\author{Jakub Jankowski\inst{1}\fnmsep\thanks{\email{Jakub.Jankowski@fuw.edu.pl}} }
\institute{Faculty of Physics, University of Warsaw, ul. Pasteura 5, 02-093 Warsaw, Poland}
\abstract{
We review classical results on holographic entanglement entropy utilizing the Hamilton-Jacobi approach. Possibility of using the entanglement entropy as a probe of confinement is shortly discussed in the context of lattice data.
} 
\maketitle

\section{Introduction}
\label{sec:intro}

Since the seminal paper of Srednicki \cite{Srednicki:1993im}
entanglement entropy became a canonical quantity to study in the context of quantum field theory, and many body systems. It attracted even more attention after Ryu and Takayanagi related it to the area of extremal surfaces in holographically dual theory \cite{Ryu:2006bv} (for a review see \cite{Nishioka:2009un,Headrick:2019eth}). Among many applications most notably Ryu-Takanayagi proposal opened a new field of research, addressing the possibility of gravity emerging from a coarse-grained description of some quantum degrees of freedom \cite{Harlow:2018fse}. To name one more interesting consequence, following directly from Ryu-Takayanagi formula, is the usage of the entanglement entropy as a measure of confinement~\cite{Klebanov:2007ws}. This possibility has been later explored in the context of lattice QCD \cite{Buividovich:2008gq}.

In this note we are going to review classical computations of holographic entanglement entropy in some $d$-dimensional field theories, that are holographically dual to certain $d+1$-dimensional theories with gravity, utilizing the Hamilton-Jacobi (HJ) equation.
We will consider two cases: one of conformal field theories dual to asypmtotically Anti de Sitter ($AdS$) spaces, and second of confining theories dual to solitonic geometries \cite{Klebanov:2007ws}.
The HJ approach allows us to directly find the area of the extremal surface, by formulating computation as a classical mechanics problem, without the necessity of finding the detailed geometric shape of the extremal surface. The idea is that, by exploring this formulation one will be able to solve problems too difficult for a direct method used up to date. The list includes for example entanglement entropy in a boost invariant flow of $\mathcal{N}=4$ super Yang-Mills \cite{Pedraza:2014moa,Bellantuono:2016tkh,Ecker:2020}, in which case the hope is that  hydrodynamic corrections can be viewed as treatable systematic perturbations. Another virtue of the HJ equation is a natural regularization scheme following from the direct dependence of the on-shell action on the boundary data. This allows for a simple reformulation of the entanglement entropy computations in cases, where original approach was unnatural and tedious, e.g., for defect configurations \cite{Janik:2015oja}. The Hamilton-Jacobi formalism has already been used in the case of holographic Wilson loops \cite{Pontello:2015yla}, as well as in holographic renormalization group \cite{deBoer:1999tgo}.

\section{Holographic computation of entanglement entropy}
\label{sec:HoloEE}

Let us start by reminding some definitions.
To define the entanglement entropy we pick up a state of a theory, say the ground state, represented by a density matrix $\rho=|0\rangle\langle0|$. We choose a spatial region $A$, and trace out all the degrees of freedom from its complement, leaving the reduced density matrix $\rho_A={\rm tr}_{A^c} \rho$. Finally, the entanglement entropy is defined as the von-Neumann entropy of $\rho_A$, i.e.,
\begin{equation}
    S_A=-{\rm tr}\left(\rho_A\log\rho_A\right)~.
\end{equation}
When one considers a theory which has a gravitational dual then the holographic prescription relates $S_A$ to the area of a a co-dimension two extremal surface, homologous to the chosen boundary region $A$. Explicitly the formula reads
\begin{equation}
    S_A=\frac{{\rm Area}(\gamma_A)}{4 G_N}~,
    \label{eq:RT}
\end{equation}
where $G_N$ is Newton's constant of the dual spacetime, and $\gamma_A$ is the surface area \cite{Ryu:2006bv}. In this way a simple geometric object encodes quantum, non-local correlations of the dual field theory, encoded in the entanglement entropy. This notion that can be extended to other quantum information quantities like complexity  \cite{Stanford:2014jda}. This story is definitely far from the end. 

\section{Asymptotically $AdS$ geometries}

In order to utilize Eq. (\ref{eq:RT}) we need to specify the geometry. We will be working in general $AdS_{d+1}/CFT_d$ setup using a general class of metric functions from Ref. \cite{Bhattacharya:2012mi}
\begin{equation}
ds^2 = g_{\mu\nu}dx^\mu dx^\nu= \frac{R^2}{z^2}\left[-f(z)dt^2+g(z)dz^2 + \sum_{i=1}^{d-2}dx_i^2\right]~,
\label{eq:geom}
\end{equation}
which includes the empty $AdS_{d+1}$ space, with $f(z)=g(z)=1$, as well as $AdS_{d+1}$ black hole, with $f(z)=1/g(z)=1-(z/z_H)^d$.
\footnote{The parameter $R$ sets the scale of the corresponding space time, and is conveniently set to $R=1$.}
The choice of homologus surfaces simply means that  $\partial\gamma_A=\partial A$, i.e., that both regions have the same boundary. To be more specific we choose the entangling region to be a semi infinite strip of length $2a$ in the $x_1=x$ direction, and infinite extent in all other directions. With this symmetry assumptions embedding is described by one symmetric function $z(x)$, where $x\in[-a,a]$. The $d-1$ dimensional metric induced on the surface is given by a standard formula
$G_{\alpha\beta}=g_{\mu\nu}\partial_\alpha x^\mu\partial_\beta x^\nu$~,
with the area determined by its volume element
\begin{equation}
    {\rm Area}(\gamma_A)=2R^{d-1}L^{d-2}\int_{-a}^{a}\frac{dx}{z^{d-1}}\sqrt{g(z)z'(x)+1}~,
    \label{eq:Area}
\end{equation}
where $L\rightarrow\infty$ is the extend in the $d-2$ spatial directions. Because entangling surface is space-like nothing depends on the $f(z)$ function, determining the $tt$ component of the metric tensor.

In order to get a well defined variational problem we need to specify boundary conditions. One, following from symmetry requires, that $z'(0)=0$. The second one comes from demanding that $\partial\gamma_A=\partial A$, which implies that $z(-a)=z(a)=0$. However, because Eq. (\ref{eq:Area}) will have divergences for $z\rightarrow0$, we will put a finite cut-off into considerations imposing that $z(-a)=z(a)=z_1$, and then considering the limit $z_1\rightarrow0$ with $a=\rm fixed$.

\section{Hamilton-Jacobi equation in the $AdS$ geometries}
\label{sec:HJEgeoms}

Now, in order to find the detailed shape of extremal surface, standard procedure is to solve the Euler-Lagrange equations following from action (\ref{eq:Area}). Then one evaluates back the action (\ref{eq:Area}) on a solution to get the extremal area.
Here, we will slightly twist that approach, treating the whole computation as one dimensional classical mechanics problem, and formulating it in terms of the Hamilton-Jacobi equation.
This allows us to directly solve for the on-shell value of the action, i.e., directly compute the desired area without explicitly finding the extremal surface.

Omitting numerical prefactors in Eq. (\ref{eq:Area}) we are facing the following Langrangian for a surface embedding
\begin{equation}
\mathcal{L}= \frac{1}{z^{d-1}}\sqrt{g(z)z'^2+1}~,
\end{equation}
with the embedding profile $z(x)$. The canonical momentum is defined to be
\begin{equation}
p = \frac{\partial\mathcal{L}}{\partial z'(x)}= \frac{g(z)z'}{z^{d-1}\sqrt{g(z)z'^2+1}}~,
\end{equation}
giving rise to the inverse relation
\begin{equation}
z'(x)=\frac{- p z^d}{\sqrt{g(z)\left(-p^2z^{2d}+g(z)z^2\right)}}~.
\end{equation}
The resulting Hamiltonian reads
\begin{equation}
\mathcal{H}(z,p,x) = \frac{\partial\mathcal{L}}{\partial z'(x)}z'(x) -\mathcal{L}(x)= -\frac{1}{z^d}\sqrt{\frac{z^2g(z)-p^2z^{2d}}{g(z)}}~.
\end{equation}
Within the boundary conditions specified at the end of previous section the on-shell action is a function of two variables $\mathcal{S}(a,z_1)$, and satisfies the Hamilton-Jacobi equation
\begin{equation}
    \partial_a\mathcal{S}+\mathcal{H}\left(z_1,\partial_{z_1}{S},a\right)=0~,
\end{equation}
where $\mathcal{H}=\mathcal{H}(z,p,x)$, which explicitly reads
\begin{equation}
 -\frac{1}{z_1^d}\sqrt{\frac{z_1^2g(z_1)-(\partial_{z_1} \mathcal{S})^2z_1^{2d}}{g(z_1)}} +\partial_a \mathcal{S} = 0~.
\end{equation}
Because the Hamiltonian is $x$-independent, we can use the following ansatz
\begin{equation}
\mathcal{S}(a,z_1) = \widetilde{S}(z_1) + E a+\mathcal{S}_0~,
\label{eq:sep}
\end{equation}
for some constant $E$. Inserting that into the HJ equation we get
\begin{equation}
\widetilde{S}(z_1) = -\int_{z_0}^{z_1} dz \sqrt{g(z)\left(z^{2-2d}+E^2\right)}~,
\end{equation}
where $z_0=z(x=0)$. Using the above, and taking the $E$ derivative of Eq. (\ref{eq:sep}) we get 
\begin{equation}
a = -\frac{\partial\widetilde{S}(z_1)}{\partial E} = E \int_{z_1}^{z_0} dz \sqrt{\frac{g(z)}{z^{2-2d}-E^2}}~.
\label{eq:a}
\end{equation}
The second integration constant is fixed by demanding that $\mathcal{S}(a=0,z_1)=0$, which results in $\mathcal{S}_0=0$. Eventually we get
\begin{eqnarray}
\mathcal{S}(a,z_1) &=& 2R^{d-1}L^{d-2}\int_{z_1}^{z_0} dz z^{2-2d}\sqrt{\frac{g(z)}{z^{2-2d}-E^2}} =\\
&=& 2R^{d-1}L^{d-2}\int_{z_1}^{z_0} dz \frac{\sqrt{g(z)}}{z^{d-1}\sqrt{1-E^2z^{2d-2}}}~,
\label{eq:intS}
\end{eqnarray}
which is exactly equation (5) from Ref. \cite{Bhattacharya:2012mi}
upon the identifications $E=1/z_*^{d-1}$, and $z_*=z(x=0)$. The above identification comes from the fact that $E$ is the value of the Hamiltonian, that can be computed at $x=0$.
This result, as promised, reproduces the right answer with a direct computation. The dependence on $a$ in Eq. (\ref{eq:intS}) is convoluted in the value of $z_0$ parameter by means of Eq.~ (\ref{eq:a}).
The divergent part of the EE is proportional to the perimeter of the entangling region, and it is a consequence of infinite number of ultraviolet degrees of freedom.

\section{Confining geometry}

Gauge/gravity duality is best understood in the case of conformal theories, that are dual to the $AdS$ space. Yet, the physically most interesting examples include systems with confinement, which can also be modelled in terms of classical geometry \cite{Witten:1998zw}. However, the precise and systematic knowledge in that case is much more difficult to grasp due to the lack of clear asymptotic properties. 

Using the entanglement entropy as a probe of confinement relies on a simple observation that if we take as a subsystem a strip of length $a$, then the entanglement entropy computed for that strip measures degrees of freedom at energy scales $\Lambda\sim1/a$. If there is a confining interaction, above a certain critical length scale $a_c$ there should be no degrees of freedom, and the EE should be "trivial". To be more precise, the EE should approach a constant for large distances, that is in the deep infrared limit of $\Lambda\rightarrow0$, or attain a constant value at a certain length scale.   

Such a behaviour can be demonstrated explicitly in a number of holographic models of confinement 
\cite{Klebanov:2007ws}. Here, we will revisit within the HJ equation an example from review article \cite{Nishioka:2009un}. The confining geometry is given by the $AdS$ solitonic solution \cite{Witten:1998zw}
\begin{equation}
    ds^2=R^2 \frac{dr^2}{r^2f(r)}+\frac{r^2}{R^2}\left(-dt^2+f(r)d\chi^2+dx_1^2+dx_2^2\right)~,
    \label{eq:geomconf}
\end{equation}
where $f(r)=1-r^4_0/r^4$, and $\chi$ is a compact direction with period $L_\chi=\pi R^2/r_0$, chosen to avoid conical singularity at $r=r_0$.
This solution is obtained from a standard $AdS$ black hole by a double Wick rotation. Dual gauge theory is $\mathcal{N}=4$ supersymmetric Yang-Mills theory on $\mathds{R}^{1,2}\times S^1$. Due to anti-periodic boundary conditions for fermins along the $\chi$ direction, fermions acquire masses at the tree level, while bosons only through the radiative corrections. This implies that supersymmetry is maximally broken.
In the low energy limit, i.e., for $E\ll 1/L_\chi$, this model shows similar behaviour as pure Yang-Mills theory in three dimensions, which is known to be confining \cite{Witten:1998zw}. In the chosen coordinate system in geometry Eq. (\ref{eq:geomconf}) asymptotic region, where dual field theory is located is at $r=\infty$.

\section{Hamilton-Jacobi equation for confining geometry}

As previously, we will be after entanglement entropy for a strip like geometry with finite extent in the $x_1=x$ direction. The Ryu-Takayanagi prescriptions works in the same way, so we have to again seek for an extremal surface homologous to the entangling region.

The novelty in this case is that in order to realize the intuition described in the previous section we need two solutions of the HJ equation with the same boundary condition. One is a connected surface, similar to the one found in the $AdS$ case, that corresponds to the deconfined state, as it has non-trivial dependence on $a$. This should be favourable solution at small distances. In contrast, for large distances, the solution should consist of two independent segments extending from the edges of our chosen strip into the dual geometry. In that case there should be no dependence on $a$, realizing confining properties of the model.

Within chosen symmetries, the generic embedding will be described by a $r(x)$ function with the following area functional
\begin{equation}
    {\rm Area}(\gamma_A)=L_\chi L^{d-2}\int_{-a}^a dx\frac{r}{R}\sqrt{r'^2+\frac{r^4f(r)}{R^4}}~,
\end{equation}
which, after removing decorations, gives the Lagrangian
\begin{equation}
    \mathcal{L}=\frac{r}{R}\sqrt{r'^2+\frac{r^4f(r)}{R^4}}~.
\end{equation}
As previously we have defined the momentum as
\begin{equation}
    p=\frac{\partial\mathcal{L}}{\partial r'(x)}=\frac{R r(x) r'(x)}{\sqrt{R^4 r'(x)^2+r(x)^4-r_0^4}}~,
\end{equation}
giving the inverse relation
\begin{equation}
    r'(x)=\frac{p \sqrt{r_0^4-r(x)^4}}{R \sqrt{p^2 R^2-r(x)^2}}~.
\end{equation}
In this case the resulting Hamiltonian is
\begin{equation}
    \mathcal{H}(r,p,x)=\frac{\partial\mathcal{L}}{\partial r'(x)}r'(x)-\mathcal{L}(x)=
    \frac{r_0^4-r(x)^4}{R^3}\sqrt{\frac{p^2 R^2- r(x)^2}{r_0^4-r(x)^4}}~.
\end{equation}
The boundary conditions are specified exactly the same way as previously giving, giving the HJ equation
\begin{equation}
    \partial_a\mathcal{S}+\mathcal{H}\left(r_1,\partial_{r_1}{S},a\right)=0~,
\end{equation}
which now depends on $\mathcal{S}=\mathcal{S}(a,r_1)$.
The explicit form is
\begin{equation}
    \partial_a\mathcal{S}+\frac{r_0^4-r(x)^4}{R^3}\sqrt{\frac{(\partial_{r_1}\mathcal{S})^2 R^2- r(x)^2}{r_0^4-r(x)^4}}=0~,
\end{equation}
As in the conformal case we can use separation of variables to solve the HJ equation
\begin{equation}
    \mathcal{S}(a,r_1)= \widetilde{S}(r_1)+E a~.
    \label{eq:Sconf}
\end{equation}
which gives an explicit solution
\begin{equation}
    \widetilde{S}(r_1)=-\int_{r_*}^{r_1}dr\frac{\sqrt{r^{10}-E^2 r^4 R^6-r_0^4 \left(2 E \sqrt{r^4-r_0^4} R^3+r_0^4-E^2 R^6\right)}}{r^4 R}~.
\end{equation}
Taking the $E$ derivative of Eq. (\ref{eq:Sconf}) we get
\begin{equation}
a = -\frac{\partial\widetilde{S}(r_1)}{\partial E} =\int_{r_*}^{r_1}dr\frac{E \left(r_0^4-r^4\right) R^5-r_0^4 \sqrt{r^4-r_0^4} R^2}{r^4 \sqrt{r^{10}-E^2 r^4 R^6-r_0^4 \left(2 E \sqrt{r^4-r_0^4} R^3-E^2 R^6+r_0^4\right)}}~.
\label{eq:aconf}
\end{equation}
Eventually we get the desired entropy
\begin{equation}
    \mathcal{S}(r_1,a)=-\int_{r_*}^{r_1}dr \frac{r^{10}-2 E^2 r^4 R^6-r_0^4 \left(3 E \sqrt{r^4-r_0^4} R^3+r_0^4-2 E^2 R^6\right)}{r^4 R \sqrt{r^{10}-E^2 r^4 R^6-r_0^4 \left(2 E \sqrt{r^4-r_0^4}
   R^3+r_0^4-E^2 R^6\right)}}~.
\end{equation}
Using the fact that Lagrangian is $x$ independent we can find the relation $E=\frac{r_*^3}{R^3}f(r_*)$.

The crucial point now is to realise that Eq. (\ref{eq:aconf}) implies that $a\leq a_{\rm crit}\sim L_\chi$, which means connected solutions only exist at small enough length scales. This means that there is a second solution competing with the connected one, i.e., the disconnected solution. It can be obtained by requiring that $\partial_a\mathcal{S}=0$, which implies 
\begin{equation}
    -R^4 r_1^2+R^6(\partial_{r_1} \mathcal{S})^2=0~,
\end{equation}
which elementally leads to
\begin{equation}
    \mathcal{S}(r_1)=\frac{r_1^2}{R^2}-\frac{r_0^2}{R^2}~.
\end{equation}
%


\begin{figure}
\begin{center}
	\includegraphics[width=0.8\textwidth]{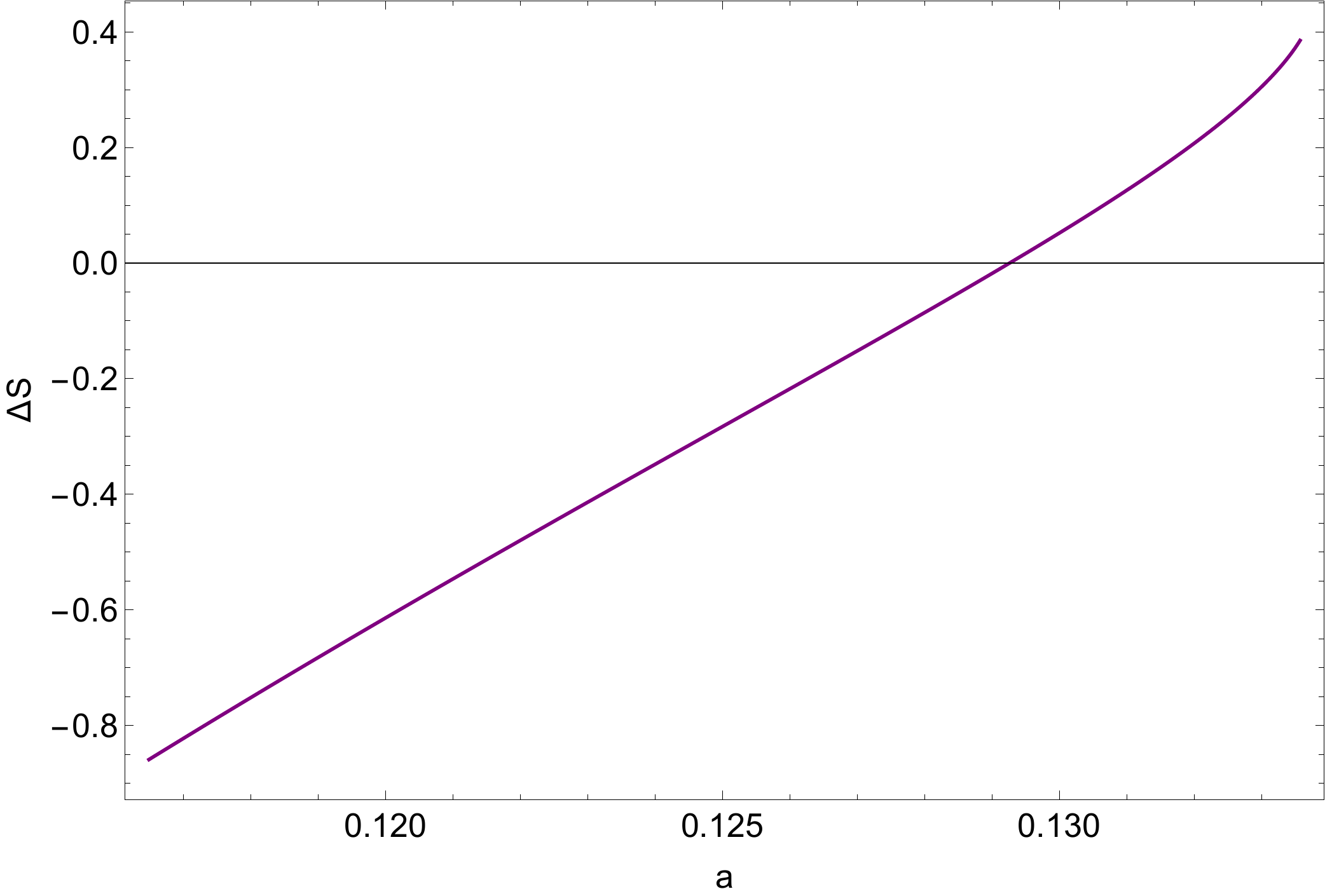}
\caption{Difference of entanglement entropy for connected and disconnected Ryu-Takayanagi surfaces. Below a certain length scale connected solution is favourable (is of smaller area), and above it disconnected solution dominates, making the entanglement trivial.}
\label{fig:holoEE}
\end{center}
\end{figure}


Now we can study the difference of entanglement entropies between connected and disconnected surfaces $\Delta S_a =S^{\rm con}(a)-S^{\rm dis}(a)$, which clearly shows that well above $a_{\rm crit}$ the connected solution is not favourable (it is not monimal).
This is illustrated in Fig. \ref{fig:holoEE}. As anticipated, above a certain length scale the entanglement entropy is constant signalling confinement. Eventually, the physical solution is a 
continous gluing of two configurations according to the sign of $\Delta S_a$.
As a consistency check let us note, that the physical solution is a concave function of $a$, as expected from the strong sub-additivity of the von Neumann entropy~\cite{Nishioka:2009un}.

\section{Meaning of holography and lattice results}

What we can learn from holography are always ballpark results, which point us towards generic behaviour that we can look for in other theories/models. This is no different in the case of confinement. Based on intuition that we have just developed we may expect, in the large-$N$ confining gauge theory, the finite part of the EE to have the following behaviour 
\begin{equation}
    S(a)|_{\rm finite}= -V F(a)~,
    \label{eq:Sexp}
\end{equation}
where $F(a)\sim c_1N^2a^{-d-2}$ for $a<a_{\rm c}$, and $F(a)\sim c_2N^2$ for $a>a_{\rm c}$ \cite{Klebanov:2007ws,Buividovich:2008yv}.
In Eq.~(\ref{eq:Sexp}) $V$ is the volume of the entangling region, and $c_1$ and $c_2$ depend on a specific theory.
Note, that in this scenario the overall scaling with $N$ does not change. Such a change in the $N$-scaling would require transition in the background geometry of the Hawking-Page type \cite{Witten:1998zw}, but that goes beyond the semi-classical approximation. Another important remark is that the EE can be computed both for vacuum configurations, as well as at finite temperatures/densities.  

In order to get a UV cut-off independent measure of effective degrees of freedom we may introduce the following function
\begin{equation}
   C(a)= \frac{a^{d-2}}{V}\frac{d S(a)}{d a}~.
\end{equation}
The expectation from a large-$N$ holographic computation is that this function should be monotonically decreasing, and exhibit a sharp vertical falloff at a certain critical length scale. For finite $N$ we expect the behaviour to be smoothed.

Surprisingly, $C(a)$ can be reliably studied in lattice gauge theories \cite{Buividovich:2008gq,Buividovich:2008yv,Rabenstein:2018bri}.
For example, 
the results for the $d=3+1$ $SU(2)$ gauge theory computed on the lattice (see figure 3 from Ref. \cite{Buividovich:2008yv}) show a rapid decrease in a number of effective degrees of freedom over the length scales $0.2\leq a\leq 0.6~$fm. A similar behaviour is seen for $SU(3)$ and $SU(4)$ gauge groups \cite{Rabenstein:2018bri}. Great aspect of quantities like $C(a)$ is that those are not restricted to cases of non-dynamical quarks, like the Polyakov loops, and in turn are a good candidate of a universal confinement order parameter. An additional feature of $C(a)$ is the sensitivity to Fermi surfaces, and fermionic collective states in general, an advantage in exploration of low temperature, and high density states. The drawback is that the entanglement entropy is very hard to compute in lattice QCD.




\section*{Acknowledgements}

I would like to thank Micha{\l} Spali{\'n}ski and Hesam Soltanpanahi Sarabi for useful discussions. 
This research was  supported by the Polish National Science Centre (NCN) grant 2016/23/D/ST2/03125.

\newpage

\end{document}